\documentclass[aps,prd,amsfonts,amsmath,twocolumn]{revtex4}

\usepackage{color}
\usepackage{graphicx}

\usepackage{epstopdf}

\newcommand {\be} {\begin{equation}}
\newcommand {\ba} {\begin{eqnarray}}
\newcommand {\ee} {\end{equation}}
\newcommand {\ea} {\end{eqnarray}}
\newcommand{\psl}{\!\not\! p}
\newcommand{\qsl}{\!\not\! q}

\begin{document}

\title{Nucleon electromagnetic and gravitational form factors from holography}

\author{Zainul Abidin and Carl E.\ Carlson}
\affiliation{
Department of Physics, College of William and Mary, Williamsburg, VA 23187-8795, USA}

\date{March 27, 2009}

\begin{abstract}
The electromagnetic form factors of nucleons are calculated using an AdS/QCD model by considering a Dirac field coupled to a vector field in the 5-dimensional AdS space.  We also calculate a gravitational or energy-momentum form factor by perturbing the metric from the static AdS solution. We consider both the hard-wall model where the AdS geometry is cutoff at $z_0$ and the soft-wall model where the geometry is smoothly cut off by a background dilaton field. 
\end{abstract}

\maketitle

%
\section{Introduction}
%

The anti de Sitter space/conformal field theory (AdS/CFT) correspondence~\cite{Maldacena:1997re,Witten:1998qj,Gubser:1998bc} is a conjecture which holds the possibility of obtaining accurate results in the strong coupling limit of gauge theories from classical calculations of gravitationally interacting fields in a higher dimensional space.   The original correspondence was between a paricular string theory in 10D and a particular conformal field theory in 4D, namely the large $N_C$ limit of ${\mathcal N} =4$ super Yang-Mills theory.

A particular implementation motivated by the original AdS/CFT correspondence is the ``bottom-up'' approach,  introduced in~\cite{Erlich:2005qh,Da Rold:2005zs},  which is a way of using the AdS/CFT correspondence as motivation for modeling QCD starting from a 5D space.    One may think that one has reached the point where the 10D string theory of the original AdS/CFT correspondence has been reduced to a gravitational theory on AdS$_5$, and one then asks what terms should exist in the Lagrangian.  The terms are chosen based on simplicity, symmetries, and relevance to the problems one wishes to study.

QCD is not a conformal field theory, so one also needs to break a corresponding symmetry in the AdS space, in order to obtain, for example, discrete hadron masses.  Two schemes which have the virtue of being analytically tractable are the hard-wall model, where the AdS space is sharply cutoff and a boundary condition imposed, and the soft-wall model where extra interactions are introduced which have an effect akin to warping the metric and suppressing long distance propagation in the fifth dimension.

Having chosen a Lagrangian and a cutoff scheme, one can study the phenomenological consequences for the 4D correspondent theory,  and compare the results to data.
Much of the work has focused on the bosonic sector~\cite{Polchinski:2001tt,Polchinski:2002jw,Brodsky:2003px,deTeramond:2005su,Brodsky:2006uqa,Brodsky:2007hb,Brodsky:2008pf,Karch:2006pv,Grigoryan:2007vg,Grigoryan:2007my,Grigoryan:2007wn,Grigoryan:2008up,Grigoryan:2008cc,Kwee:2007dd,Kwee:2007nq, BoschiFilho:2005yh,BallonBayona:2007vp,Abidin:2008ku,Abidin:2008hn}.   The works include studies of spin-1 vector and axial states, pseudoscalars, and glueballs.   Masses, decay constants, and charge radii that can be compared to experimental data agree with experimental data at the roughly 10\% level.

Studying fermions with the AdS/CFT correspondence is technically more complicated than studying bosons.  Two approaches have been pursued.  One approach is to follow upon the bosonic studies, and treat the fermions as Skyrmions within the model~\cite{Hata:2007mb,Pomarol:2007kr,Pomarol:2008aa}.  The other approach is to begin with a theory in the 5D sector that has fundamental fermion fields interacting with an AdS gravitational background~\cite{Henningson:1998cd,Mueck:1998iz,Contino:2004vy,Hong:2006ta,Brodsky:2008pg}.  One can also consider a hybrid of the two approaches, where one begins with fermions as Skyrmions of a 5D model, and uses the Skyrmion model to obtain parameters and interaction terms of another 5D Lagrangian where the fermion fields appear as explicit degrees of freedom~\cite{Hong:2007kx,Hong:2007ay,Hong:2007dq}.

We here pursue the AdS/CFT correspondence within a model where the fermion stands as an explicit field in the 5D Lagrangian.  We will particularly be interested in obtaining results for the electromagnetic and gravitational form factors.  For the electromagnetic form factors,  there is already work reported in the literature, particularly for the hard-wall model, and we will quote some results from this material, adding some useful detail.  The derivation of the tensor or gravitational form factors is new.

One point in mapping fermion fields from a 5D theory to a 4D theory is that not all components of the fermion spinor are independent.  In both theories, for massive fermions, one can obtain the (ill-named) right handed part of the field from the left handed part of the field, or \textit{vice-versa}.  An early 4D discussion of this is in~\cite{Brown:1958zz}.  Thus, as one begins by finding exact solutions for fermions in an AdS background, one can only choose boundary conditions for the independent components, which one can choose to be the left handed ones.  The left handed 5D fermions on the boundary are sources for right handed fermionic currents in the 4D theory, and the corresponding left handed fermion currents can either be obtained from these, or can be consistently obtained from the derived right handed fermions in AdS space.

When we study the soft wall model for fermions, the usual procedure of producing a ``soft wall'' by inserting an interaction with a background dilation field via an overall exponential factor does not by itself lead to normalizable solutions.  (Indeed, it is possible to remove the overall exponential factor in the fermion case by rescaling the field.)  We will introduce an additional interaction with the background field by using an analog of a scalar potential, that is, by adding a dilation interaction to the mass term.  The resulting equations describing the interaction with both AdS gravity and the soft-wall potential can be solved exactly in the classical limit, with the normalizable solutions having the feature common to soft-wall models that their functional dependence in the extra dimension contains a generalized Laguerre polynomial.

Our presentation will focus on the soft-wall model.  It is easy to switch our equations to the hard-wall model just by dropping the soft-wall interaction term and using a suitable boundary condition.  We will comment on this and show hard-wall as well as soft-wall results.   We will obtain expressions for both electromagnetic and gravitational form factors, and present the result for the nucleon radii as well as for general momentum transfer.  We will find that radii obtained from the gravitational momentum form factor is smaller than radii obtained electromagnetically, as is also the case for the meson sector~\cite{Abidin:2008ku,Abidin:2008hn}.

The model, focusing on the baryonic degrees of freedom, is outlined in Sec.~\ref{sec:model},  and the two-point functions are worked out in Sec.~\ref{sec:twopoint}.   The form factors, both electromagnetic and gravitational are discussed in Sec.~\ref{sec:ff}, and a summary is given in Sec.~\ref{sec:summary}.

%
\section{The Model}           \label{sec:model}
%

In a $d$-dimensional field theory, the generating function is given by
 \be
 \mathcal{Z}_{CFT}[\Phi^0]=\left< \exp{\left(i\int d^dx  \mathcal{O}(x) \Phi^0(x)\right)}\right>,
 \ee
The  precise statement of the AdS/CFT correspondence is that the generating function of a $d$-dimensional CFT is equal to the partition function of a field theory in ${\rm AdS}_{d+1}$
 \be
 Z_{CFT}[\Phi^0]=e^{iS_{AdS}(\Phi_{cl})}.
 \ee
On the right hand side of the above equation, $S_{AdS}(\Phi_{cl})$ is the classical action evaluated on the solution of the equation of motion with boundary condition
\be
\lim_{z \to 0} \Phi_{cl} (x,z)= z^\Delta \Phi^0(x).
\ee
The constant $\Delta$ depends on the nature of operator $\mathcal{O}$.

The metric of  $d+1$-dimensional AdS space is given by
\be
ds^2=g_{MN}dx^M dx^N=\frac{1}{z^2}\left(\eta_{\mu\nu} dx^\mu dx^\nu - dz^2 \right), \label{metric}
\ee
where $\eta_{\mu\nu}=diag(1,-1,-1,-1)$, $\mu,\nu=0,1,2 ,...,d-1$ and we will set $d=4$. The $z$ variable extend from $\varepsilon \to 0$, which is called the UV-boundary, to $\infty$, which is the IR-boundary. In order to simulates confinement, one can use a hard-wall model \cite{Erlich:2005qh,Da Rold:2005zs,deTeramond:2005su,Grigoryan:2007vg}, by cutting off the AdS geometry at $z_0$. The mass spectrum is found to be approximately linear at large mass, $m_n \sim n$.   Alternatively,  one can use the soft-wall model \cite{Karch:2006pv,Grigoryan:2007my},  where the geometry is smoothly cut-off by a background dilaton field $\Phi(z)$. A choice for the dilaton field solution, $\Phi(z)=\kappa^2 z^2$, gives the mass spectrum that is in agreement with Regge trajectory, $m^2_n \sim n$.

Consider a Dirac field coupled to a vector field in the $d+1$-dimensional AdS space with the following action 
\ba
S_F&=&\int d^{d+1} x \sqrt{g}\, e^{-\Phi(z)} \bigg( \frac{i}{2} \bar{\Psi} e^N_A \Gamma^A D_N \Psi	 \nonumber \\
 	    &&-\frac{i}{2}(D_N \Psi)^\dagger \Gamma^0 e^N_A \Gamma^A \Psi-(M+\Phi(z))\bar{\Psi}\Psi\bigg),        \label{fermionAction}
\ea
 where for the AdS space, $e^N_A=z \delta^N_A$ is the inverse vielbein. Covariant derivative $D_N=\partial_N +\frac{1}{8}\omega_{NAB}[\Gamma^A,\Gamma^B]-iV_N$ ensure that the action satisfies gauge invariance and diffeomorphism invariance. The non-vanishing components of the spin connection are $\omega_{\mu z\nu}=-\omega_{\mu\nu z}=\frac{1}{z}\eta_{\mu\nu}$. The Dirac gamma matrices have been defined in such a way that they satisfy anti-commutation relation $\{\Gamma^A,\Gamma^B\}=2\eta^{AB}$, that is for $d=4$, we have $\Gamma^A=(\gamma^\mu,-i\gamma^5)$.  We implement the soft-wall model by adding $\Phi(z)=\kappa^2 z^2$ to the mass term. Both the Dirac and the vector fields have an $U(2)$ isospin structure. In particular, $V_N=\frac{1}{2} V^s_N +V^a_N t^a$, where $t^a$ is an $SU(2)$ generator normalized by Tr$(t^a t^b)=\delta^{ab}/2$.
 
The Dirac field satisfies the following equation of motion
\be
\big[i e^N_A \Gamma^A D_N -\frac{i}{2}(\partial_N\Phi)\, e^N_A \Gamma^A - (M+\Phi(z))\big]\Psi=0.
\ee
Evaluating the action on the solution, we obtain
\ba
S_F[\Psi_{cl}]&=& \int d^d x\, \left(-\frac{i}{2}\sqrt{g}e^{-\kappa^2 z^2} \bar{\Psi} z \Gamma^z \Psi \right)_\epsilon^{z_{IR}}      \nonumber\\
&=& \int d^d x\,\frac{-1}{2z^d}e^{-\kappa^2 z^2} \left(\bar \Psi_L \Psi_R - \bar \Psi_R \Psi_L\right)\Big|_\epsilon^{z_{IR}},     \ \ \ \ 
\ea
where $\Psi_{R,L}=(1/2)(1\pm \gamma^5)\Psi$. For hard-wall model  the IR boundary is located at finite $z_{IR}=z_0$,  while for the soft-wall model the $z$ variable extends to infinity, {\it i.e.}, $z_{IR}=\infty$. In the case of hard-wall model, the IR boundary term can be removed by requiring that either $\Psi_L(z_{IR})=0$ or $\Psi_R(z_{IR})=0$.  

Following \cite{Henningson:1998cd,Mueck:1998iz,Contino:2004vy}, we add extra term in the UV-boundary
 \be
 \frac{1}{2}\int d^d x \sqrt{-g^{(d)}} \left(\bar{\Psi}_L \Psi_R +\bar{\Psi}_R \Psi_L\right)_{\varepsilon}.
 \ee 
This term preserves the $O(d+1,1)$ isometry group of the original action and does not change the equation of motion. The action becomes
\be
S_F=\int d^d x \left(\frac{1}{z^d}\bar{\Psi}_L \Psi_R \right)_{\varepsilon} \label{SFsurf}
\ee

The Dirac field $\Psi_{R,L}$ in momentum space can be written in terms of a product of  $d$-dimensional boundary fields $\Psi^0_{R,L}$  and profile functions or  the bulk-to-boundary propagators $f_{R,L}$,  {\it i.e.}, $\Psi_{R,L}(p,z)=z^\Delta f_{R,L}(p,z) \Psi^0(p)_{R,L}$,  where $p$ is the momentum in $d$-dimensions. We choose $\Psi^0_L(p)$ as the independent source field which corresponds to the spin-$\frac{1}{2}$ baryon operator $\mathcal{O}_R$ in the $d$-dimensional field theory. Hence, $\Delta$ is chosen such that the equation of motion allows $f_L(p,\varepsilon)=1$. 

The left handed and the right handed components of the spin-$\frac{1}{2}$ field operators in d-dimensional flat space are not independent, since they are related by the Dirac equation. This is realized by the relationship between $\Psi^0_L$ and $\Psi^0_R$, chosen to satisfies  $\psl \Psi^0_R(p)=p \Psi^0_L(p)$. Ignoring the interaction term with the vector field, the equation of motion for the Dirac field becomes (for $\Phi = \kappa^2 z^2$),
\ba
\left(\partial_z - \frac{d/2+M-\Delta+2\Phi}{z}\right)f_R&=&-pf_L,\nonumber\\
\left(\partial_z - \frac{d/2-M-\Delta}{z}\right)f_L&=&pf_R, \label{fLfR}
\ea
where $p\equiv\sqrt{p^2}$.

In addition to (\ref{fermionAction}) or (\ref{SFsurf}), we have the kinetic term of the vector field
 \be S_V=\int d^{d+1} x~ e^{-\Phi} \sqrt{g}  {\rm Tr}\left(-\frac{F^2_V}{2g^2_5}\right)   ,
 \ee
where $F^V_{MN}=\partial_M V_N-\partial_N V_M$.The transverse part of the vector field can be written as $V_\mu(p,z)=V(p,z)V^0_\mu(p)$.  At the UV-boundary, the bulk-to-boundary propagator satisfies $V(p,\varepsilon)=1$.  According to the AdS/CFT dictionary, the $V^0_\mu (p)$ is the source for the 4D current operator $J^V_\mu$. The equation of motion in the $V_z=0$ gauge is given by \cite{Grigoryan:2007my}
\be
\left[ \partial_z\left(\frac{e^{-\Phi}}{z}\partial_z\right) +\frac{e^{-\Phi}}{z}p^2\right] V(p,z)=0. \label{vectoreom}
\ee
The normalizable mode,  is a solution of the above equation with eigenvalue $p^2=M^2_n$ which  corresponds to the mass of the $n$-th Kaluza Klein mode of vector meson \cite{Erlich:2005qh}.  For the soft-wall model \cite{Grigoryan:2007my}, the mass eigenvalues are $M^2_n=4\kappa^2(n+1)$, where $n=0,1,\ldots$.  For the hard-wall model the mass eigenvalues are $M_n=\gamma_{0,n+1}/z_0$, where $\gamma_{0,n+1}$ is the $n+1$-th zeros of the Bessel function $J_0$.

%
\section{Two-point function}   				\label{sec:twopoint}
%

\subsection{Soft-wall Model}
To have $f_L(p,\varepsilon)= 1$ for $\Phi(0)=0$ and $f_R$ not singular requires $\Delta=\frac{d}{2}-M$. The equation of motions of the profile functions (\ref{fLfR}), with $\Phi(z)=\kappa^2 z^2$, become
\begin{align}
 &\bigg[\partial^2_z-\frac{2\left(M+\kappa^2 z^2\right)}{z}\partial_z +\frac{2\left(M-\kappa^2 z^2\right)}{z^2} + p^2\bigg]f_R=0\,,\nonumber\\
&\bigg[\partial^2_z-\frac{2\left(M+\kappa^2 z^2\right)}{z}\partial_z + p^2\bigg]f_L=0.\label{dfLfR}
\end{align}
The general solution is given by Kummer's functions of the first and the second kind. Requiring that the profile functions vanish at infinity, we obtain
\ba
f_L(p,z)&=&N_L \,  U\left(-\frac{p^2}{4\kappa^2}, \frac{1}{2}-M; \xi\right),\\
f_R(p,z)&=&N_R \, \xi^{\frac{1}{2}} U\left(1-\frac{p^2}{4\kappa^2}, \frac{3}{2}-M; \xi\right),
\ea
where $\xi=\kappa^2 z^2$.  From the UV-boundary condition we obtain
\be
N_L=\frac{\Gamma \left(\alpha -\frac{p^2}{4\kappa^2}\right)}{\Gamma\left(\alpha\right)},
\ee
where $\alpha=M+\frac{1}{2}$ and we have $N_R=N_L \frac{p}{2\kappa}$  from Eq.~(\ref{fLfR}).

The normalizable wave function $\psi^{(n)}_{L,R}$ for the $n$-th Kaluza Klein mode can be obtained from Eq.~(\ref{dfLfR}), by requiring that $p^2=m_n^2$. One find that solutions exist in terms of Laguerre polynomial when $m_n^2=4\kappa^2(n+\alpha)$ 
\begin{align}
&\psi^{(n)}_L(z)=n_L{\xi^{\alpha}} 
			L^{(\alpha)}_n(\xi),\\
&\psi^{(n)}_R(z)=n_R {\xi^{\alpha-\frac{1}{2}}}
			L^{(\alpha-1)}_n(\xi).
\end{align}
Imposing the normalization condition,
\be
 \int dz\, \frac{e^{-\kappa^2 z^2}}{z^{2M}} \psi^{(n)}_{L}\psi^{(m)}_{L}=\delta_{nm}\,, \label{norm}
\ee
one obtains the normalization factors, 
\ba
n_L&=&\frac{1}{\kappa^{\alpha-1}}\sqrt{\frac{2\Gamma(n+1)}{\Gamma(\alpha+n+1)}},\\
n_R&=&n_L\sqrt{\alpha+n},
\ea
For the right handed  wave function $\psi^{(n)}_R$, the normalization factor  can be obtained either by using Eq.~(\ref{fLfR}) or by imposing the above normalization condition. 

For the time-like region $p^2>0$, the profile functions have infinite number of poles which correspond to the tower of infinite Kaluza-Klein mode. To show this, we write the profile functions in different forms utilizing the Kummer transformation (\cite{AS}, p.505), 
\begin{align}
&f_L(p,z)=N_L \,  \xi^{\alpha} U\left(\alpha-\frac{p^2}{4\kappa^2}, \alpha+1; \xi\right),\\
&f_R(p,z)=N_R \, \xi^{\alpha-\frac{1}{2}} U\left(\alpha-\frac{p^2}{4\kappa^2}, \alpha; \xi\right).
\end{align}
The Kummer function can be written in integral representations, 
\ba
\hskip -2mm f_L(p,z)=\frac{\xi^{\alpha}}{\Gamma\left(\alpha \right)}\int_0^1 dx \frac{x^{\alpha+a-1}}{(1-x)^{\alpha+1}} \exp\left(-\frac{x}{1-x}\xi \right),
\ea
where $a=-p^2/(4\kappa^2)$. The integrand contains generating function for Laguerre polynomial (\cite{AS}, p.784)
\be
\frac{1}{(1-x)^{\alpha+1}}\exp\left(-\frac{x}{1-x}\xi\right)=\sum_{n=0}^{\infty} L^{(\alpha)}_n(\xi) x^n.
\ee
Performing the integrals,  one obtains
\be
f_L(p,z)=\frac{-4\kappa^2\xi^{\alpha}}{\Gamma\left(\alpha\right)}\sum_{n=0}^{\infty} \frac{L^{\left(\alpha\right)}_n(\xi)}{p^2-4\kappa^2(n+\alpha)},
\ee
which show that the mass square of the $n$-th Kaluza-Klein mode is $4\kappa^2(n+\alpha)$, as expected. Similar expansion for the right handed profile function yields
\be
f_R(p,z)=\frac{-2\kappa\, p\, \xi^{\alpha-\frac{1}{2}}}{\Gamma\left(\alpha\right)}\sum_{n=0}^{\infty} \frac{L^{\left(\alpha-1\right)}_n(\xi)}{p^2-4\kappa^2(n+\alpha)}  \, .
\ee
Notice that the Laguerre polynomials which appears in the expansion precisely match the normalizable modes. Defining $f_n= 2\kappa/(\Gamma(\alpha)\, n_R)=\psi^{(n)}_R(\varepsilon)/\varepsilon^{2M}$, the profile functions become
\ba
f_L(p,z)&=&\sum_{n=0}^\infty \frac{f_n m_n\, \psi^{(n)}_L(z)}{p^2-m_n^2},\label{fLExpand}\\
f_R(p,z)&=&\sum_{n=0}^\infty \frac{f_n p\, \psi^{(n)}_R(z)}{p^2-m_n^2}. \label{fRExpand}
\ea

The 5D fermion action (\ref{SFsurf}) can now be written in terms of sum over modes
\ba
S_F&=&\int \frac{d^d p}{(2\pi)^d} \bar{\Psi}^0_L(p) \frac{f_R(p,\varepsilon)\psl}{\varepsilon^{2M}p}\Psi^0_L(p) \nonumber\\
&=&\sum_n \int \frac{d^d p}{(2\pi)^d}\bar{\Psi}^0_L(p)\left(\frac{-f_n^2 P_R \psl}{p^2-m_n^2}\right) \Psi^0_L(p),
\ea
where $P_R=(1+\gamma^5)/2$ is the right handed chirality projector. From the AdS/CFT correspondence and the appropriate functional derivatives, we have 
\be
\int d^d x \, e^{iqx} \left<0\left|\mathcal{T}\mathcal{O}_R(x)\bar{\mathcal{O}}_R(0) \right|0\right>=\sum_{n}\frac{if_n^2 P_R \qsl}{q^2-m_n^2}   \,.
\ee
One may also define the decay constant $f_n$ from $\left<0\left|\mathcal{O}_R(0)\right|p\right>=f_n u_R(p)$ and obtain the same result by inserting a set of intermediate states. In order to obtain the complete two-point function $\left<\mathcal{O}\bar{\mathcal{O}}\right>$, one also needs the left handed chirality operator $\mathcal{O}_L$, which can be obtained from the right handed one using $\mathcal{O}_L(p)=(\psl/p)\mathcal{O}_R(p)$.

\subsection{Hard-wall Model}
For the hard-wall model $\kappa=0$, and IR boundary is at $z_{IR}=z_0$.  The mass eigenvalue of the Kaluza Klein mode depends on which propagator vanish at IR boundary. We will set $f_R(z_0)=0$, such that there is no massless mode. Requiring that $f_L(p,\varepsilon)=1$, the solution can be written in terms of Bessel function
\ba
f_L&=&\frac{\pi}{\Gamma(\alpha)}\left(\frac{pz}{2}\right)^{\alpha} \left(\frac{Y_{\alpha-1}(pz_0)}{J_{\alpha-1}(pz_0)}J_{\alpha}(pz)-Y_{\alpha}(pz)\right),\\
f_R&=&\frac{\pi}{\Gamma(\alpha)}\left(\frac{pz}{2}\right)^{\alpha} \left(\frac{Y_{\alpha-1}(pz_0)}{J_{\alpha-1}(pz_0)}J_{\alpha-1}(pz)-Y_{\alpha-1}(pz)\right).\nonumber
\ea

The normalizable mode, again, setting $p^2=m_n^2$ on Eq.~ (\ref{dfLfR}) with Dirichlet boundary condition $\phi^{(n)}_R(z_0)=0$ and $\phi^{(n)}_L(\varepsilon)=0$ gives
\ba
\phi^{(n)}_L(z)&=&\frac{\sqrt{2} z^{\alpha} J_{\alpha}(m_n z)}{z_0 J_{\alpha}(m_n z_0)}  , \nonumber\\
\phi^{(n)}_R(z)&=&\frac{\sqrt{2} z^{\alpha} J_{\alpha-1}(m_n z)}{z_0 J_{\alpha}(m_n z_0)} .
\ea
Both satisfy normalization condition given in Eq.~(\ref{norm}). The mass eigenvalue determined by $J_{\alpha-1}(m_n z_0)=0$. One can easily see that the location of the pole of the profile function $f_{R,L}(p,z)$, in the time-like region $p^2>0$, are precisely at the mass eigenstates $m_n^2$. 

As in the soft-wall model, the bulk-to-boundary propagators can be written in terms of sum over normalizable modes given in Eq. (\ref{fLExpand}) and (\ref{fRExpand}),
where for the hard-wall model
\be
f_n=\frac{\sqrt{2} (\frac{m_n}{2})^{\alpha-1}}{\Gamma(\alpha) z_0 J_{\alpha}(m_n z_0)}.
\ee

%
\section{Form factors}					\label{sec:ff}
%
\subsection{Electromagnetic Form Factors}
For spin-$\frac{1}{2}$ particles, the electromagnetic current matrix element  can be written in terms of two independent form factors
\ba
&&\left<p_2,s_2\big| {J}^\mu(0) \big|p_1,s_1\right>=\\
&&\qquad u(p_2,s_2)\left(F_1(Q)\gamma^\mu + F_2(Q) \frac{i\sigma^{\mu\nu}q_\nu}{2m_n}\right)u(p_1,s_1),\nonumber
\ea
where $q=p_2-p_1$ and $Q^2=-q^2$. In this paper,  our interest is on the electromagnetic current operator of nucleons which can be written in terms of isoscalar and isovector currents
\be
J^\mu_{p,n}= \chi_i \left(\frac{1}{2}J^\mu_S \delta_{ij}+ {J^a_V}^\mu t^a_{ij}\right)\chi_j,
\ee
where $\chi=(1,0)$ for proton, and $\chi=(0,1)$ for neutron. According to the AdS/CFT dictionary, the 4D isoscalar $J^\mu_S$ and isovector  ${J^a_V}^\mu$ current operators correspond to the isoscalar and isovector part of the 5D gauge field respectively. In the simplest model, the profile function of both  satisfy Eq. (\ref{vectoreom}). 

The isovector matrix element can be extracted from the 3-point function by the following relation
\ba
&&\lim_{p^0_{1,2}\to E^n_{1,2}} (p_1^2-m_n^2)(p_2^2-m_n^2)\int d^4 x d^4 y \nonumber\\
&&\qquad\times e^{i(p_2 x -qy-p_1 w)}\left<0\big|\mathcal{T} \mathcal{O}^i_R(x) {J^a}^\mu(y) \bar{\mathcal{O}}^j_R(w)\big|0\right>\nonumber\\
&&=f_n^2\chi^i u(p_2,s_2)\left<p_2,s_2\big|{J^a}^\mu(0)\big|p_1,s_1\right>\bar{u}(p_1,s_1)\chi^j\nonumber\\
&&\qquad \times\delta^{(4)}(p_2-p_1-q)\,,
\ea
and analogously for the isoscalar current.

Relevant term in the action (\ref{fermionAction}) which contribute to the 3-point function is given by    
\be
S_{F}=\int d^5 x \, \sqrt{g} e^{-\Phi} \bar{\Psi} e^M_A \Gamma^A V_M \Psi.
\ee
However the above term only provides the $F_1$ form factor. Hence, we should add the following gauge invariant term to the action
\be
\eta_{S,V}\int d^5 x \, \sqrt{g} e^{-\Phi} i\frac{1}{2} \bar{\Psi}\, e^M_A\, e^N_B \,[\Gamma^A,\Gamma^B]\, F^{(S,V)}_{MN} \Psi. \label{Addterm}
\ee
We shall use different parameters:  $\eta_S$ for the isoscalar component and $\eta_V$ for the isovector component of the vector field. They are fixed by the experimental values of the proton and the neutron magnetic moments.  

Defining invariant functions
\begin{align}
&C_1(Q)=\int dz\, e^{-\Phi}\frac{V(Q,z)}{2 z^{2M}}\left({\psi_L}^2(z)+{\psi_R}^2(z)\right)\label{C1}\,,\\
&C_2(Q)=\int dz\, e^{-\Phi}\frac{\partial_z V(Q,z)}{2 z^{2M-1}}\left({\psi_L}^2(z)-{\psi_R}^2(z)\right),\label{C2}\\
&C_3(Q)=\int dz\, e^{-\Phi}\frac{2 m_n V(Q,z)}{z^{2M-1}} \psi_L(z)\psi_R(z), \label{C3}
\end{align}
where $Q^2=-q^2>0$, we obtain the electromagnetic form factors for the proton
\ba
F^{(P)}_1(Q)&=&C_1(Q)+\eta_{P}C_2(Q),\\
F^{(P)}_2(Q)&=&\eta_{P}C_3(Q). 
\ea
For the neutron, the $F_1$ and the $F_2$ form factors are solely come from  (\ref{Addterm}). We have 
\ba
F^{(N)}_1(Q)&=&\eta_{N}C_2(Q),\\
F^{(N)}_2(Q)&=&\eta_{N}C_3(Q),
\ea
with parameters $\eta_P$ and $\eta_N$ are defined by $\eta_{P,N}=(\eta_V \pm \eta_S)/2$. 

\begin{figure}[t]

\begin{center}
\includegraphics[width = 8.0 cm]{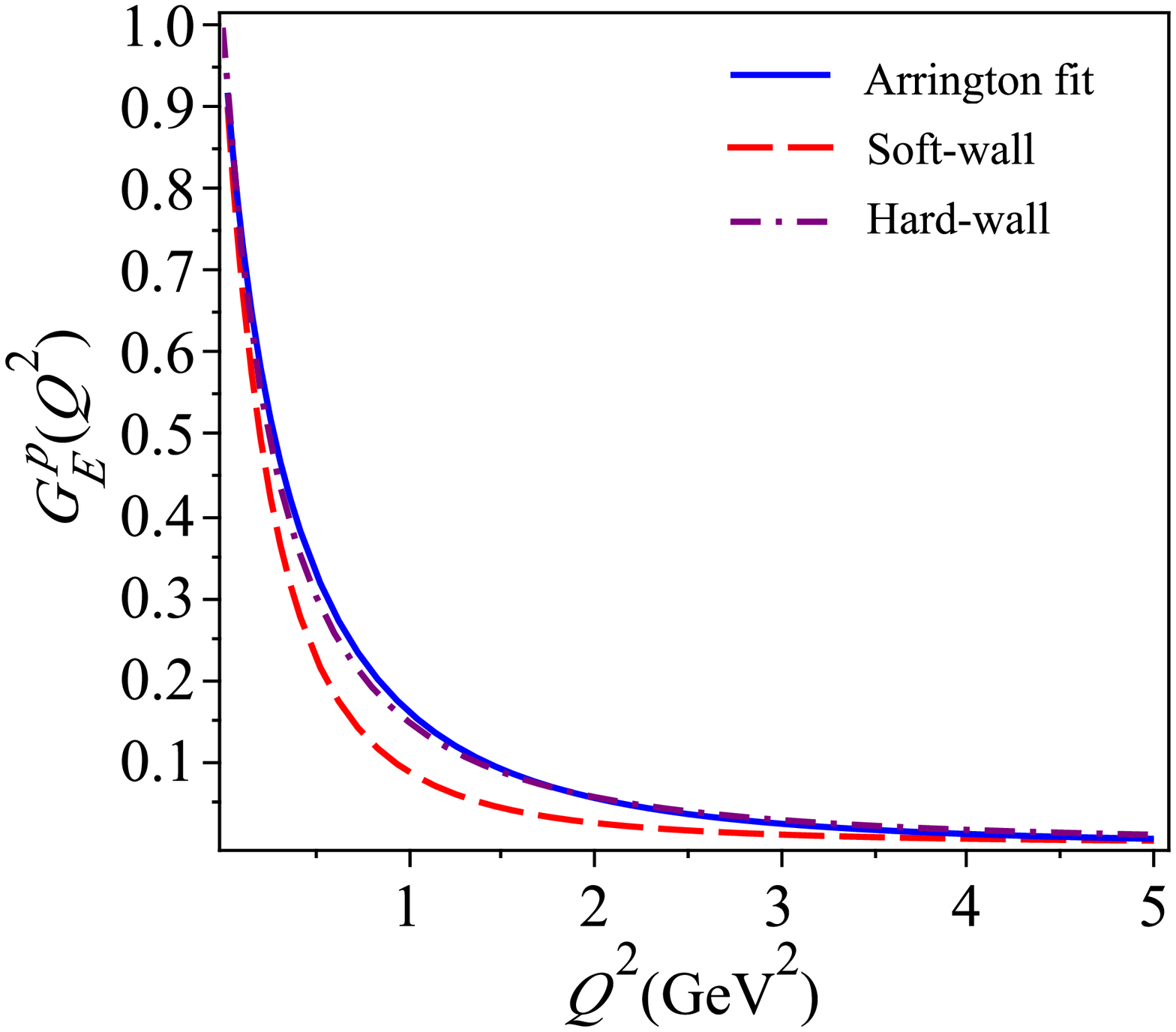} 
\includegraphics[width = 8.0 cm]{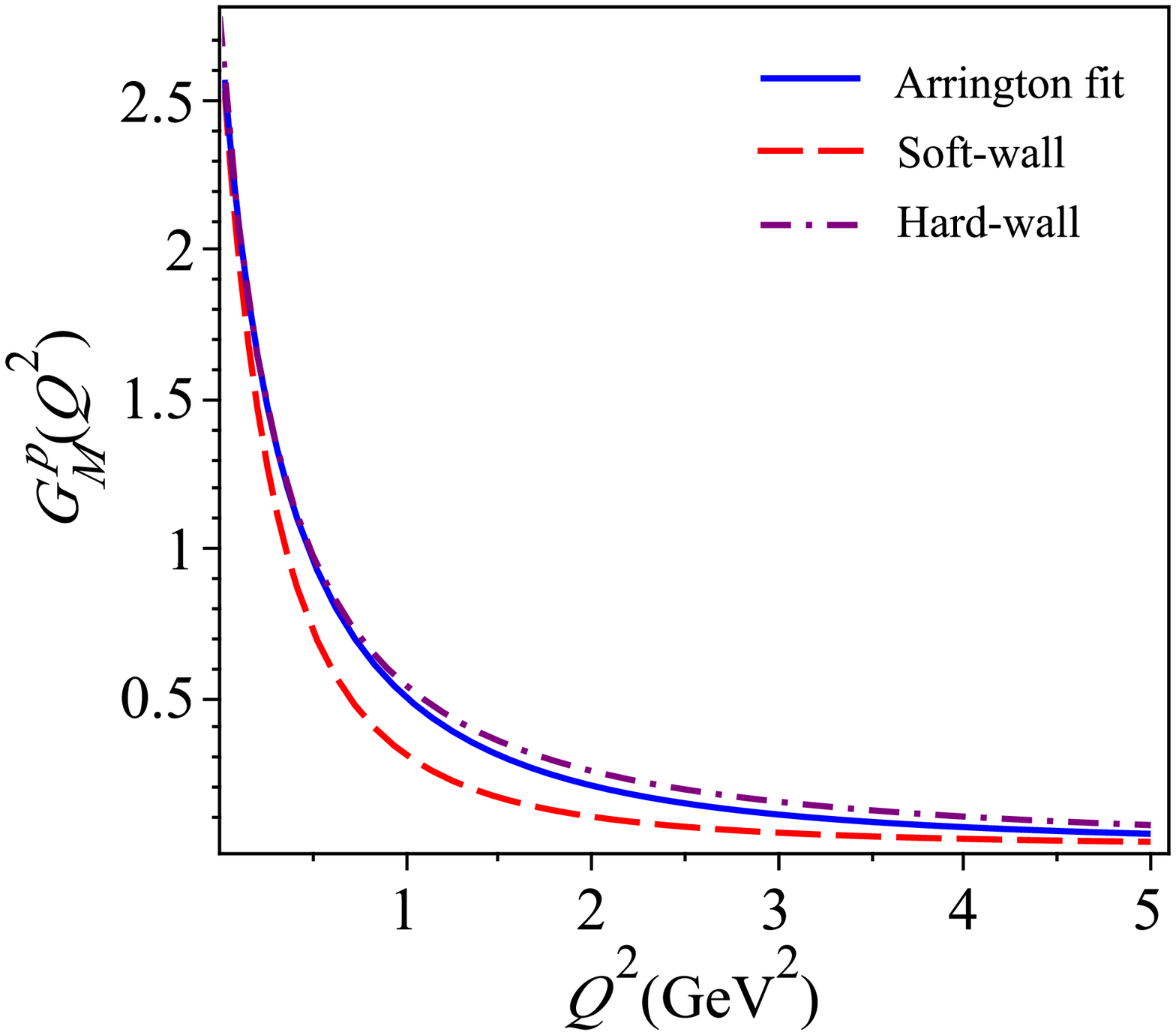} 
\end{center}

\vglue -8 mm
\caption{The red dashed line and the purple dot-dashed line are the electromagnetic form factors of proton from  the soft-wall and the hard-wall model of AdS/QCD respectively. The solid blue line is the corresponding form factor from the Arrington empirical fit \cite{Arrington:2007ux}}
\label{fig:1plots}
\end{figure}

{\it In the soft-wall model}, the bulk-to-boundary propagator of the vector field is given by\cite{Grigoryan:2007my}
\ba
V(Q,z)&=&\Gamma\left(1+a\right) U\left(a,0;\xi\right)\nonumber\\
&=& a\int_0^1 dx\, x^{a-1} \exp\left(-\frac{x}{1-x}\xi\right)    \,.
\ea
where again $a=Q^2/(4\kappa^2)$. Integral in Eq.(\ref{C1}-\ref{C3}) can be evaluated analytically.  For the lowest state $n=0$,  we obtain
\ba
C_1(Q)&=&\frac{\frac{1}{2}\left(2+\frac{a}{\alpha+1}\right)}{\left(\frac{a}{\alpha+1}+1\right)\left(\frac{a}{\alpha}+1\right)...\left(a+1\right)},\\
C_2(Q)&=&\frac{-a(1-\alpha a)\left(\alpha+2\right)^{-1}\left(\alpha+1\right)^{-1}}{\left[\left(\frac{a}{\alpha+2}+1\right)...\left(a+1\right)\right]},\\
C_3(Q)&=& \frac{4\alpha }{\left(\frac{a}{\alpha+1}+1\right)\left(\frac{a}{\alpha}+1\right)...\left(a+1\right)}.
\ea
One can check that $F^{(P)}_1(0)=1$ and $F^{(N)}_1(0)=0$. 

In the limit of large momentum transverse, the electromagnetic form factors for the proton becomes
\ba
 F^{(P)}_1(Q)&=&\frac{\alpha ! (2\kappa)^{2\alpha}}{2 Q^{2\alpha}}\left(1+2\eta_P\, \alpha\right),\\
 F^{(P)}_2(Q)&=&\frac{4\alpha (\alpha+1)! \,(2\kappa)^{2\alpha+2}}{Q^{2\alpha + 2}}.
\ea 
Hence, $M=\frac{3}{2}$ which corresponds to $\alpha=2$ gives the correct large momentum scaling. The constant $\kappa$, was simultaneously fixed to the proton's and the $\rho$-meson's mass. The best fit, given by $\kappa=0.350\,{\rm GeV}$, gives the proton's mass $0.990\,{\rm GeV}$ and $\rho$-meson's mass $0.700\,{\rm GeV}$.

Parameters $\eta_P$ and $\eta_N$ can be determined by matching the value of $F_2(0)$ with the experimental data: $1.793$ for proton and $-1.913$ for neutron. One obtain, for $\alpha=2$, $\eta_P=0.224$ and $\eta_N=-0.239$.

The charge radius for the proton is defined by 
\ba
\left<r_C^2\right>_{p}=-\frac{6}{G_E(0)}\frac{d G_E(0)}{d Q^2}   \,,
\ea
where $G_E(Q)=F_1(Q)-Q^2F_2(Q)/(4m_p^2)$. One obtains
\ba
\left<r_C^2\right>_{p}=\frac{5}{2\kappa^2}+\frac{\eta_P}{8\kappa^2}+\frac{6 F^{(P)}_2(0)}{4 m_p^2}=(0.961\, {\rm fm})^2,
\ea
which, in terms of rms-radius, is about $10$ percent larger than the experimental result $\left<r_C\right>=(0.877\, {\rm fm})$. 

For the neutron, the charge radius is defined  by
\ba
\left<r_C^2\right>_{n}=-6\frac{d G_E(0)}{d Q^2}.
\ea
One obtains
\ba
\left<r_C^2\right>_{n}=\frac{\eta_N}{8\kappa^2}+\frac{6 F^{(N)}_2(0)}{4 m_p^2}=(-0.136\, {\rm fm}^2),
\ea
which is an acceptably well result compared to the experiment $\left<r^2_C\right>=(-0.112\, {\rm fm}^2)$.

\begin{figure}[t]

\begin{center}
\includegraphics[width = 8.0 cm]{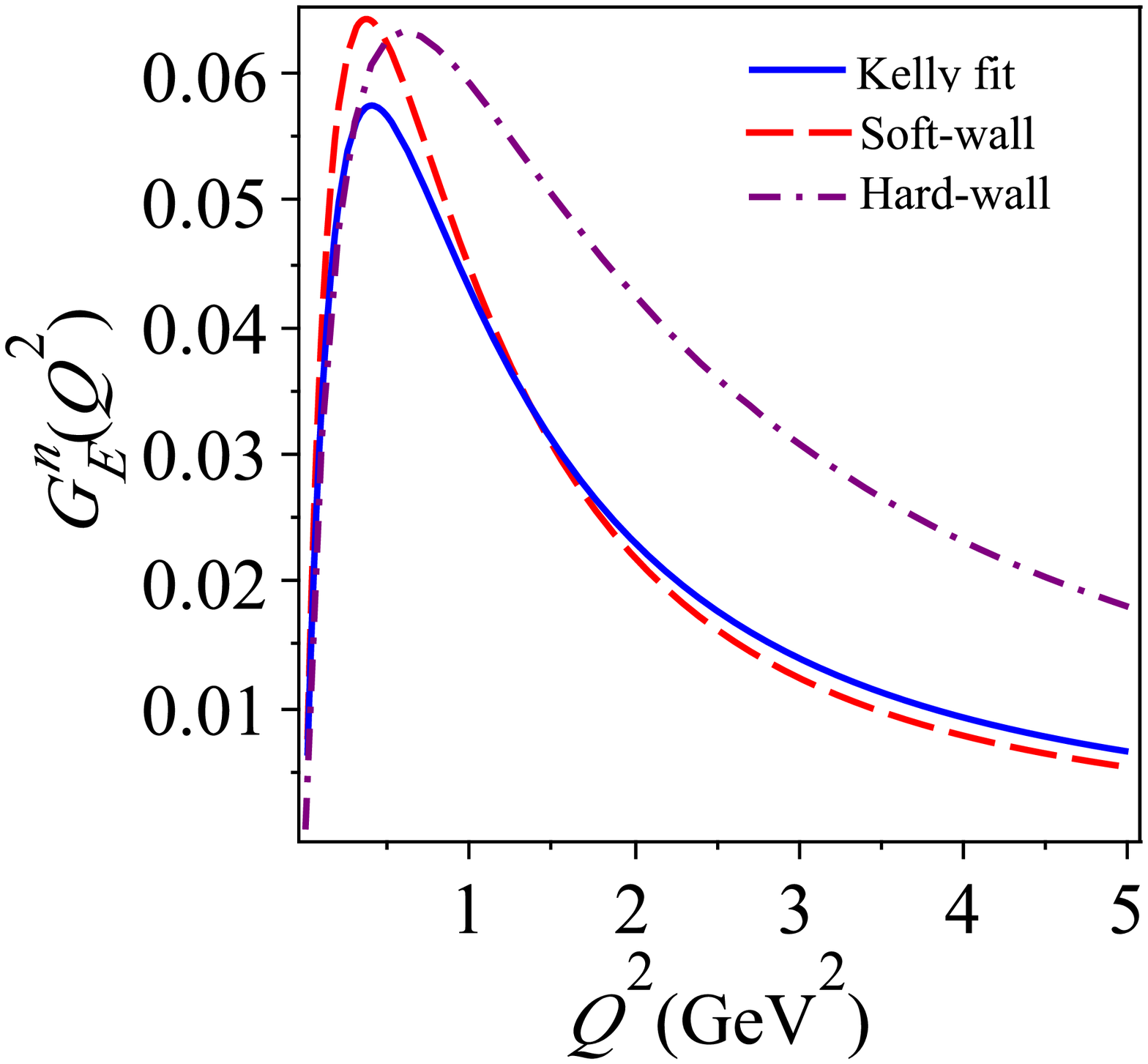}
\includegraphics[width = 8.0 cm]{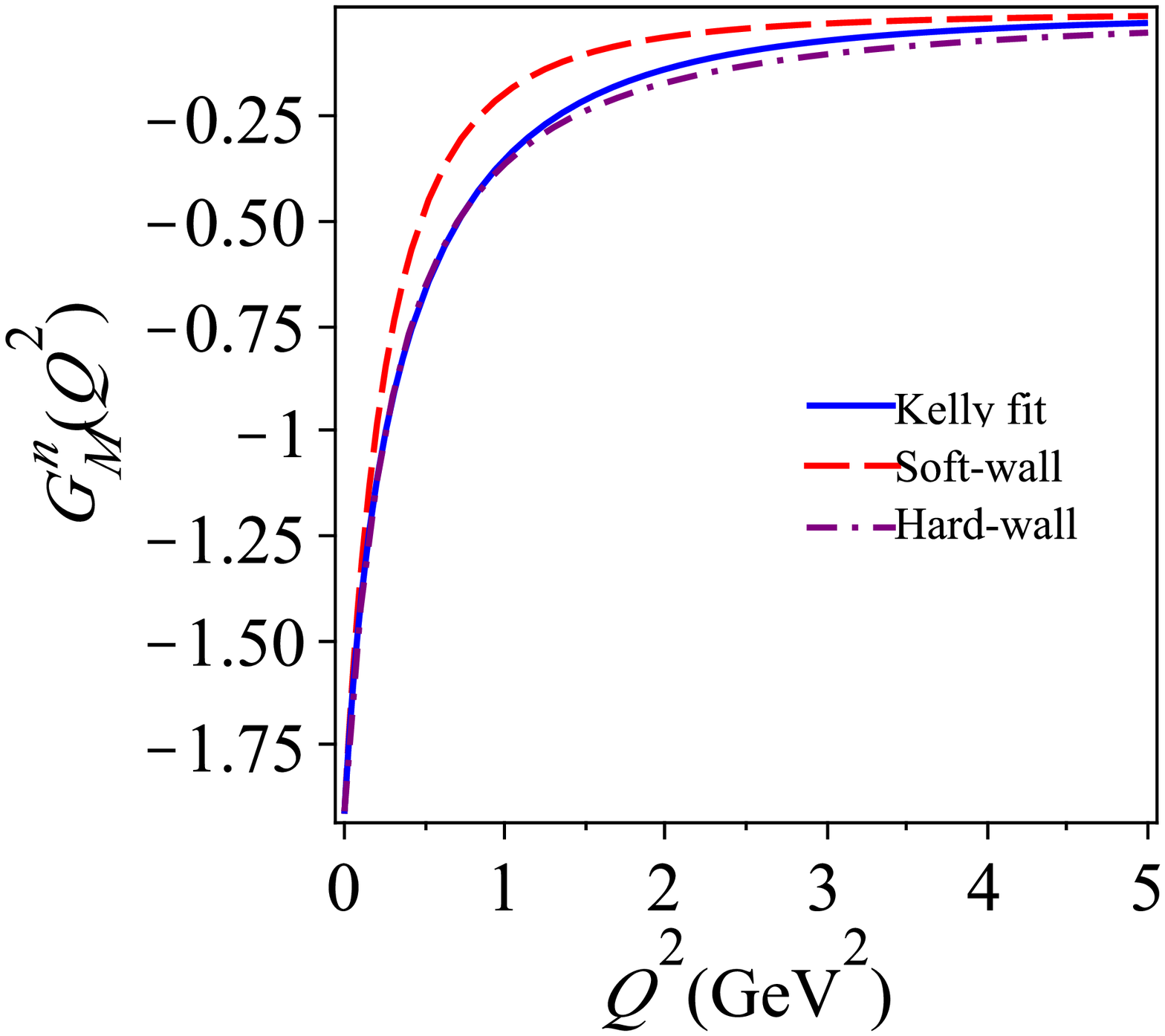} 
\end{center}

\vglue -8 mm
\caption{The red dashed line and the purple dot-dashed line are the electromagnetic form factors of neutron from  the soft-wall and the hard-wall model of AdS/QCD respectively. The solid blue line is the corresponding form factor from the Kelly empirical fit \cite{Kelly:2004hm}}
\label{fig:2plots}
\end{figure}

{\it For the hard-wall model}, the bulk-to-boundary propagator is given by \cite{Grigoryan:2007vg}
\be
V(Q,z)=Qz\left(\frac{K_0(Qz_0)}{I_0(Qz_0)}I_1(Qz)+K_1(Qz)\right).
\ee
The parameter $z_0$ determines both the mass of the nucleon and $\rho$-meson.  We set $z_0=(0.245\, {\rm GeV})^{-1}$, which fits the measured proton's mass. 

In the large $Q^2$ region, $V(Q,z)\to Qz K_1(Qz)$, which behaves like exponential. It has significant support near $z=0$ only. Therefore, one can replace $\phi_R^2(z)\pm\phi_L^2(z)$ by its approximate form near $\varepsilon$, that is, $\pm f_n^2 z^{4\alpha-2}$ . One obtains
\ba
C_1(Q)&=& \frac{f_n^2}{2Q^{2\alpha}}\int_0^\infty dw\, w^{2\alpha} K_1(w),\\
C_2(Q)&=&\frac{f_n^2}{2Q^{2\alpha}}\int_0^\infty dw\, w^{2\alpha+1} K_0(w),\\
C_3(Q)&=& \frac{2 m_n^2\, f_n^2}{\alpha Q^{2\alpha+2}}\int_0^\infty dw\, w^{2\alpha+2} K_1(w),
\ea
where the integral can be solved analytically to obtain
\ba
C_1(Q)&=&2^{\alpha-2}\alpha ! (\alpha-1)! \frac{f_n^2}{Q^{2\alpha}},\\
C_2(Q)&=&2^{\alpha-1}(\alpha !)^2 \frac{f_n^2}{Q^{2\alpha}},\\
C_3(Q)&=&2^{\alpha+2}(\alpha-1)! (\alpha+1)!\,  m_n^2\,\frac{ f_n^2}{ Q^{2\alpha+2}}
\ea 
Just as in the soft-wall model,  the $F_1$ form factor falls off correctly like $Q^{-4}$, when $\alpha=2$. Fixing the $F^{(P)}_2(0)$ to the experimental value $1.793$, one obtains $\eta_P=0.448$. Hence, for the proton
\be
F^{(P)}_1(Q)= \frac{3.64}{Q^4},\qquad F^{(P)}_2(Q)= \frac{12.37}{Q^6} \,.\\
\ee
For the neutron, fixing $F^{(N)}_2(0)$ to the experimental value $-1.913$, we have  $\eta_N=-0.478$.

In the limit where $Q^2\to0$, the bulk-to-boundary propagator of the vector field can be expanded as
\be
V(Q,z)=1-\frac{Q^2z^2}{4}\left(1-2\ln\left(\frac{z}{z_0}\right)\right), \label{Vq0}
\ee 
hence, in this limit,
\be
\partial_z V(Q,z)=Q^2z\ln\left(\frac{z}{z_0}\right).\label{dVq0}
\ee
Substituting Eq.(\ref{Vq0}) and Eq.(\ref{dVq0})  to Eq.(\ref{C1}-\ref{C3}) and taking the derivative with respect to $Q^2$, we obtain the Dirac  radius for the proton $\left<r^2_1\right>_p=(0.843~{\rm fm})^2$, which corresponds to the charge radius $\left<r^2_C\right>_p=(0.910~{\rm fm})^2$. For the neutron, we obtain $\left<r^2_C\right>_n=(-0.125~{\rm fm}^2)$. These calculated charge radius are in better agreement with experimental results compared to the soft-wall model.

In Fig.~\ref{fig:1plots} we show plots of $G_E$ and $G_M$ form factors using AdS/QCD model and compare it with empirical fit given in \cite{Arrington:2007ux}. Figure~\ref{fig:2plots} shows the corresponding plots for the neutron with the empirical fit given in~\cite{Kelly:2004hm}.

\subsection{Gravitational Form Factors}
The most general structure of stress tensor matrix element for spin-$\frac{1}{2}$ particles can be written in terms of three form factors
\ba
&&\left<p_2,s_2\big| {T}^{\mu\nu}(0) \big|p_1,s_1\right>=u(p_2,s_2)\bigg(A(Q)\gamma^{(\mu} p^{\nu)}\\
&& \quad + B(Q) \frac{ip^{(\mu} \sigma^{\nu)\alpha}q_\alpha}{2m_n}+C(Q)\frac{q^\mu q^\nu-q^2 \eta^{\mu\nu}}{m}\bigg)u(p_1,s_1),\nonumber
\ea
where $p=(p_1+p_2)/2$. This matrix element can be extracted from the following 3-point function
\be
\left<0\big|\mathcal{T} \mathcal{O}^i_R(x) T^{\mu\nu}(y) \bar{\mathcal{O}}^j_R(w)\big|0\right>\,. \label{T3pf}
\ee
Stress tensor operator in 4D strongly coupled theory correspond to the metric perturbation in the bulk. 

Consider a gravity-dilaton-tachyon action \cite{Batell:2008zm,Batell:2008me}, in addition to (\ref{fermionAction}).  The metric is perturb from its static solution according to $\eta_{\mu\nu}\to\eta_{\mu\nu}+h_{\mu\nu}$. The action in the second order perturbation becomes 
\be
S_{GR}=-\int d^5 x\, \frac{e^{-2\kappa^2 z^2}}{4z^3}  \left(h_{\mu\nu,z}{h^{\mu\nu}}_{,z}+h_{\mu\nu}\Box h^{\mu\nu} \right)
\ee
where the transverse-traceless gauge conditions, $\partial^\mu h_{\mu\nu}=0$, and $h^\mu_\mu=0$, have been imposed. The profile function of the metric perturbation satisfies the following linearized Einstein equation
\be
\left[ \partial_z\left(\frac{e^{-2\kappa^2 z^2}}{z^3}\partial_z\right) +\frac{e^{-2\kappa^2 z^2}}{z^3}p^2\right] h(p,z)=0,
\ee

For the soft-wall model, the non-normalizable solution  is given by
\ba
H(Q,z)&=&\Gamma(a'+2) U(a',-1;2\xi),\\
&=&a'(a'+1) \int_0^1 dx \, x^{a'-1} (1-x) \exp\left(\frac{-2\xi x}{1-x}\right),\nonumber
\ea
where $H(Q,z)\equiv h(q^2=-Q^2,z)$ and $a'=a/2$. It satisfies $H(p,\varepsilon)=1$ and vanishes at infinity. For the hard-wall model, imposing Neumann boundary condition $\partial_z H(p,z_0)=0$, we have \cite{Abidin:2008ku}  
\be
H(Q,z)=\frac{(Qz)^2}{2}\left(\frac{K_1(Qz_0)}{I_1(Qz_0)}I_2(Qz)+K_2(Qz)\right).
\ee

In order to calculate (\ref{T3pf}), we will need terms in the 5D action in the form of $h\bar\Psi\Psi$.  The vielbeins are modified according to ${e^\mu}_\alpha \to {e^\mu}_\alpha-z h^\mu_\alpha/2$. In the transverse-traceless gauge, the determinant of the metric is unchanged from the static solution.  It can be shown that the following factor in the 5D action (\ref{fermionAction}) is unchanged under perturbation
\be
\frac{1}{8} {e^M}_C\,\Gamma^C\omega_{MAB} [\Gamma^A, \Gamma^B]\,.
\ee
Hence, remaining terms in the 5D action (\ref{fermionAction}) relevant in calculating (\ref{T3pf}) are
\be
S^{(G)}_F= \int \frac{d^5 x}{z^5} \left(\frac{-iz h_{\mu\nu}}{4}\right) \left(\bar\Psi \Gamma^\mu \tensor{\partial}^\nu \Psi\right)\,.
\ee

\begin{figure}[t]

\begin{center}
\includegraphics[width = 8.0 cm]{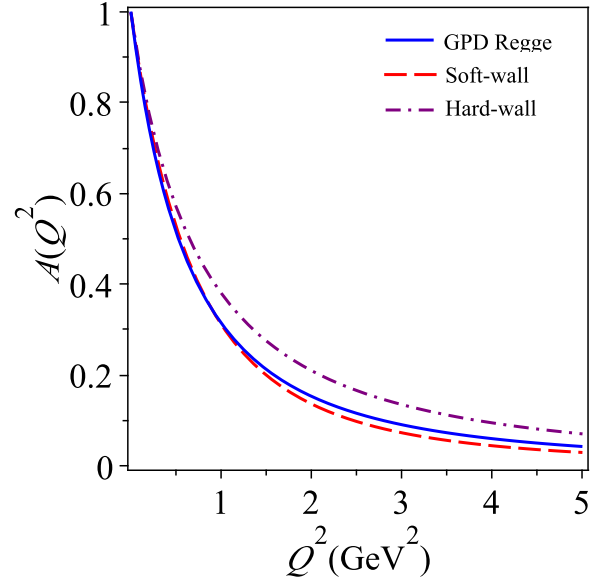}
\end{center}

\vglue -8 mm
\caption{The red dashed line is the gravitational form factor from the soft-wall model, while the solid blue line is the corresponding form factor from the integral of a GPD model~\cite{Guidal:2004nd}, and the purple dot-dash line is for the hard-wall model.}
\label{fig:3plots}
\end{figure}

Fourier transforming the fields 
\ba
S^{(G)}_F&=&\int \frac{d z}{z^{2M}} e^{-\kappa^2 z^2} \int \frac{d^4 p_2 d^4 q d^4 p_1}{(2\pi)^{12}} \nonumber\\
&&\times(2\pi)^4 \delta^4(p_2-q-p_1)\,\bar\Psi^0_L(p_2)h^0_{\mu\nu}(q) H(q,z)\nonumber\\
&&\times\frac{-1}{2}\bigg(f_L(p_2,z)f_L(p_1,z)\gamma^\mu p^\nu\nonumber\\
&&+f_R(p_2,z)f_R(p_1,z)\frac{\!\not\!p_2}{p_2}\gamma^\mu p^\nu\frac{\!\not\!p_1}{p_1} \bigg)\Psi_L(p_1)\,, \label{GravitationalAction}
\ea
where $H(Q,z)$ is the bulk-to-boundary propagator defined by $h_{\mu\nu}(q,z)=H(q,z)h^0_{\mu\nu}(q)$ and $h^0_{\mu\nu}(q)$ acts as a source for the 4D stress tensor operator.

Lorentz structure of Eq.(\ref{GravitationalAction}) shows that only $A$ form factor present. We obtain
\be
A(Q)=\int dz\, \frac{e^{-\kappa^2 z^2}}{2z^{2M}}H(Q,z)\left(\psi_L^2(z)+\psi_R^2(z)\right)  .
\ee
For the soft-wall model, an analytical solution can be obtained. In particular, for $n=0$
\ba
A(Q^2)&=&(a'+1)\bigg[-\left(1+a'+2a'^2\right)\nonumber\\
&&\qquad+2\left({a'}+2{a'}^3\right)\Phi(-1,1,a')\bigg]  ,
\ea
where $\Phi(-1,1,a')$ is the LerchPhi function.  Results are shown graphically in Fig.~\ref{fig:3plots} for both the hard-wall and soft-wall models, and compared to results obtained by integrating a model for the nucleon GPDs~\cite{Guidal:2004nd}.

The corresponding gravitational radius is
\ba
\left<r_G^2\right>&=&-\frac{6}{A(0)}\frac{d A(0)}{d Q^2}\nonumber\\
&=&\frac{3 \ln(2)}{2\kappa^2}=(0.575\, {\rm fm})^2,
\ea
which is slightly smaller than the gravitational radius obtained from the second moment integral of the modified Regge GPD model, {\it i.e.}, $0.608\, {\rm fm}$, and notably smaller than the proton charge radius.


\section{Summary}					\label{sec:summary}


We have studied baryon form factors using the AdS/QCD correspondence, and have modeled the baryons using fundamental fermions in the extra dimensional theory. We have given results for both the soft-wall and hard-wall models for both electromagnetic form factors and for the gravitational form factor $A(Q^2)$, the momentum form factor.

The soft-wall model has extra interactions whose effect is to effectively cut off propagation as one gets deeply into the extra dimension.  Originally, the soft-wall exponential modifications were simply inserted~\cite{Karch:2006pv} in order to obtain an excited hadron spectrum more in accord with observation, but it has been shown~\cite{Batell:2008zm,Batell:2008me} how to obtain the exponential factors in a  dynamical model including kinetic terms and a scalar potential for explicit dilaton and tachyon degrees of freedom.  We have followed the latter implementation here, noting that it leads to different numerical coefficients in the argument of the exponential for the vector and graviton sectors.  For the baryon sector, we implemented the soft-wall model by including also a harmonic oscillator-like scalar potential added to the mass term~\cite{Batell:2008me}.

In the bottom-up approach to modeling QCD via 5D theories and the AdS/CFT correspondence, the terms in the 5D Lagrangian are chosen based on simplicity, symmetries, and relevance to the quantities under study.  However, the most simple vector-fermion interaction yields only a Dirac form factor, so a Pauli term must be introduced in the 5D action.  This means that the overall normalization of the $F_2$ form factors is not determined \textit{ab initio}, but the shape of the form factors is fixed.

Our results for the form factors were presented both algebraically and graphically over some $Q^2$ range, with the radii corresponding to each form factor given explicitly.  In all cases, radii measured from gravitational form factors are smaller that radii measured from electromagnetic form factors.  This accords with similar observations from lattice gauge theory~\cite{Hagler:2007xi}, and one may attribute it to the fact that higher momentum fraction matter, quarks or gluons, is more heavily weighted in the momentum sum rule, and high momentum fraction partons tend to have a narrower transverse size distribution~\cite{Burkardt:2002ks}.


\begin{acknowledgments}

We thank Josh Erlich, Kostas Orginos, and Hovhannes Grigoryan  for helpful comments, and thank the National Science Foundation for support under Grant No. PHY-0555600.

\end{acknowledgments}


\end{document}